# LOW-VELOCITY IMPACTS ON PVDF TARGETS USING A LIGHT GAS GUN.

J. A. Carmona, M. Cook, J. Schmoke, R. Laufer, L. S. Matthews and T. W. Hyde, Center for Astrophysics, Space Physics and Engineering Research (CASPER), Baylor University, Waco, TX 76798-7310, USA. (Truell_Hyde@baylor.edu)

**Introduction:** Orbital debris is a constraint on the long-term health of any spacecraft and must be considered during mission planning. Varying mechanisms have been proposed to quantify the problem. Accurate in-situ data is essential with various types of sensors designed to detect orbital debris impacts employed on space missions since the 1950's [1]. The earliest of these was the PZT (piezoelectric lead zirconate titanate) sensor which was often used in-situ to measure the momentum of a particle at the time of impact. More recently, PVDF (Polyvinylidene fluoride) [2] has been employed as it exhibits piezoelectric capabilities along with the advantages of ruggedness, no bias requirement, ease of large area sensor construction, high counting rate capability, and space reliability, making it an ideal space debris sensor. Its large sensing surface area and ease of integration into a PZT sensor system makes it a desirable element in any in-situ space debris sensor.

**Procedure:** CASPER's Light Gas Gun (LGG) [3] uses helium to propel various size impactors down an instrumented beam line. For this experiment, 3/32" aluminum and chromed-steel particles were employed. A sheet of PVDF material was clamped uniformly within two aluminum frames (of dimension 5" x 5") as shown in Figure 1 and acted as the sensor element.

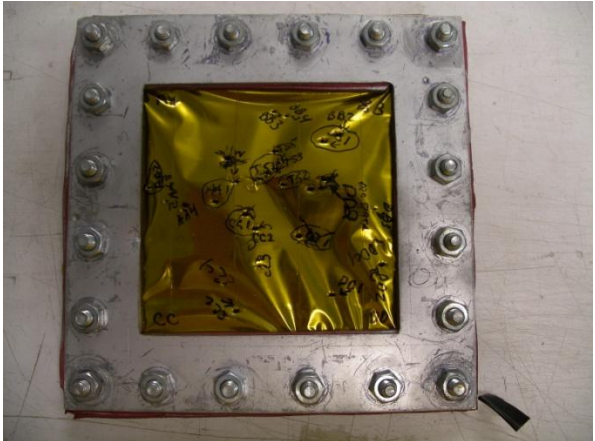

Fig. 1. PVDF sensor in its holding frame after test firing. Impact locations are marked.

A second impact (witness) plate was placed behind the target as shown in Figure 2. This witness plate was instrumented employing a PZT sensor (with the PZT donated by Dr. Harry D. Shirey, Piezo- Kinetics). Two laser / photodetector diagnostic arrays located within the beamline measure the speed of the projectile immediately upon leaving the barrel of the gun. These signals were then sent to a Tektronix 3032 oscilloscope. Data collected employing this configuration allows for the measurement of the impactor's velocity both before and after impact.

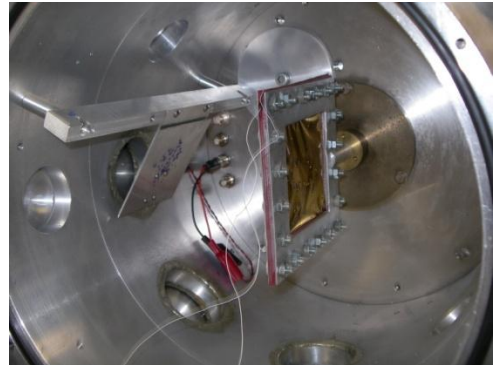

Fig. 2. PVDF and witness plate mounted within the LGG impact chamber.

The data collected in this experiment is shown in Table 1. Both the initial measured speed of the impactor the speed measured between the PVDF and PZT sensors are given along with the initial helium gas pressure used to obtain these speeds.

| Shot # | VEL (LF1-LF2) | VEL PVDF-PZT | Pressure |
|---|---|---|---|
| AA1 | 416 m/s | 303 m/s | 700 psi |
| BB1 | 404 m/s | 330 m/s | 700 psi |
| CC1 | 401 m/s | 222 m/s | 700 psi |
| DD1 | 413 m/s | 187 m/s | 700 psi |
| AA2 | 548 m/s | 439 m/s | 1200 psi |
| BB2 | 547 m/s | 454 m/s | 1200 psi |
| CC2 | 556 m/s | 450 m/s | 1200 psi |
| DD2 | 546 m/s | 448 m/s | 1200 psi |
| AA3 | 288 m/s | 168 m/s | 1500 psi |
| BB3 | 291 m/s | 200 m/s | 1500 psi |
| CC3 | 280 m/s | 190 m/s | 1500 psi |
| DD3 | 290 m/s | 192 m/s | 1500 psi |
| AA4 | 225 m/s | 123 m/s | 750 psi |
| BB4 | 238 m/s | 135 m/s | 750 psi |
| CC4 | 239 m/s | 134 m/s | 750 psi |
| DD4 | 238 m/s | 111 m/s | 750 psi |

Table 1. Data from LGG experiments.

As it can be seen, there is a change in the speed of the impactor as it passes through the PVDF material. This allows measurement of the impactor momentum transferred to the PVDF and subsequent degradation of the sensor due to damage. The sensor was divided into four quadrants with each quadrant labeled in an alphabetic pattern running from AA to DD. Each quadrant was impacted four times in order to measure signal degradation over time.

**Results:** LabVIEW was used to collect and record all experimental data with the results for a series using 3/32" aluminum projectiles shown in Figure 3 as a representative diagnostics screen shot. The upper waveform shows the signal from the PVDF material with the bottom showing the signal coming from the PZT. The remaining two signals are those produced by beamline diagnostics.

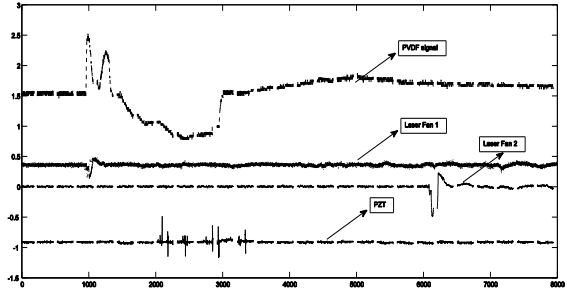

Fig. 3. Representative screen shot showing data collected from beamline diagnostics, the PVDF sensor and the PZT witness plate.

As can be seen in Figures 3 and 4, the magnitude of the initial series of shots at each of the four quadrants comprising the PVDF sensor are comparable to one another, both in maximum peak height and in time duration response. When compared to the final series of shots, shown in Figure 5, degradation of the signal appears to be minimal. The one anomaly can be seen in the BB4 series where a noticeable reduction in the peak signal amplitude is observed. This is assumed to have been created due to the close proximity between the BB1 and BB4 series.

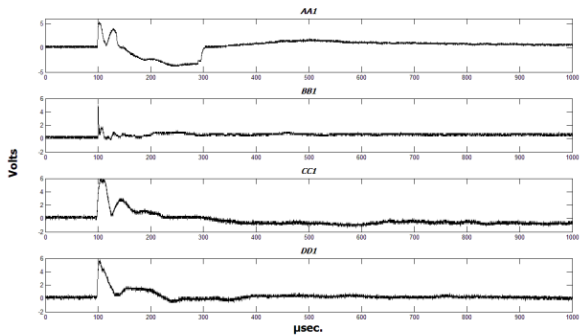

Fig. 4. Data produced by the PVDF sensor from the initial four series.

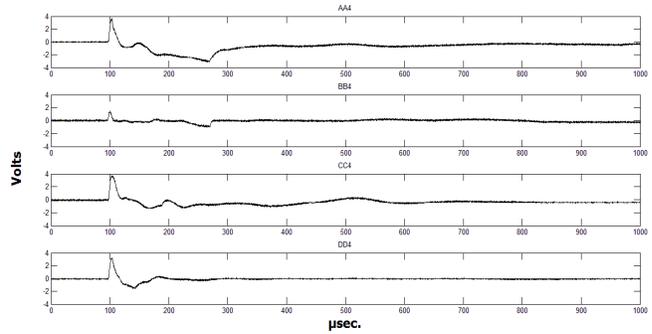

Fig. 5. Data collected from the PVDF sensor for the final four series.

**Conclusions:** As expected, the signal from the PVDF sensor proved to be both reliable and repeatable, essential conditions for any in-situ sensor. Additional PVDF data analysis (for example, employing wavelet transforms and FFTs) is necessary for a complete understanding of the data; however it is apparent that the combination of PVDF and PZT sensors provide a viable alternative for space debris detection. The CASPER LGG also continues to prove itself to be a valuable tool for both scientific research, e. g. in advanced spacecraft shielding [4], calibration for particle tracking devices [5], and in educational outreach.

Due to an agreement signed in 2007 between Baylor University and the Universitaet Stuttgart, CASPER and the Institute of Space Systems (IRS) are now collaborating in the field of space research and space technology applications. As part of this collaborative agreement CASPER is under consideration to contribute an instrument in the field of plasma and/or dust detector research to the scientific payload. The work above signals the beginning of this research and development.

**References:** [1] Alexander, W.M. *et al.* 1965. *Science,* **149**, pp. 3689-3695. [2] http://www.meas-spec.com/piezo-film-sensors.aspx [3] Carmona, J. A., *et al.* (2004) *LPS XXXV,* Abstract #1019. [4] Carmona, J. A., *et al.* (2006) *LPS XXXVII*, Abstract #1394 [5] http://www.starvisiontech.com